\documentclass[prl,twocolumn,showpacs,superscriptaddress]{revtex4}
\usepackage{graphicx}
\usepackage{color}
\usepackage{pifont}
\usepackage{amstext}   
\usepackage{amsmath}
\usepackage{graphics,amssymb}

\begin{document}

\title{ Evolution of the Fermi Surface across a Magnetic Order-Disorder Transition in the Two-Dimensional Kondo Lattice Model: A Dynamical Cluster Approach }

\author{L. C. Martin and F. F. Assaad}
\affiliation{ Institut f\"ur Theoretische Physik und Astrophysik,
Universit\"at W\"urzburg, Am Hubland, D-97074 W\"urzburg, Germany }
\begin{abstract}
We use the dynamical cluster approximation, with a quantum Monte Carlo cluster solver on 
clusters of up to 16 orbitals, to   
investigate the evolution of the Fermi surface across the magnetic order-disorder 
transition  in the two-dimensional doped Kondo lattice model. 
In the paramagnetic phase we observe the generic hybridized heavy fermion  band structure 
with  large Luttinger  volume.  
In the antiferromagnetic phase, the heavy fermion band drops below the Fermi surface 
giving  way to hole pockets  centered around  
$ {\mathbf k} = ( \pi/2 , \pi/2 ) $ and equivalent points.  In this phase Kondo screening 
does not break down but the topology of the resulting 
Fermi surface is that of a spin-density wave approximation in which the localized spins are frozen.   
\end{abstract}


\maketitle

In its simplest form, the Kondo lattice model (KLM) describes a lattice of spin 1/2 
magnetic moments coupled antiferromagnetically via an exchange coupling $J$ 
to a single band of conduction electrons and is 
believed to capture the physics of heavy fermion materials such as CeCu$_6$. The huge mass 
renormalization can be attributed to the coherent superposition of individual 
Kondo screening clouds and the resulting metallic state is characterized by a Fermi 
surface (FS) with Luttinger volume containing both conduction and localized electrons. 
At constant density, the $J$ dependence of the inverse effective mass --- or the coherence 
temperature --- has been argued to track the single ion Kondo scale \cite{Georges00,Assaad04a}. 
Competing with Kondo screening, the localized spins interact indirectly via magnetic polarization 
of the conduction band. This RKKY scale dominates at low values of the exchange coupling and  is the
driving force for the observed magnetic order-disorder quantum phase transitions in those materials
\cite{Doniach77}.
The nature of this phase transition is of current interest  following experimental results 
suggesting a sudden change in the FS topology at the quantum critical point (QCP) 
for the heavy fermion metal 
YbRh$_{2}$Si$_{2}$ \cite{Paschen04}. Driving this system from the non-magnetic heavy fermion 
metallic phase to the antiferromagnetic (AF) metallic phase causes a rapid change in the low 
temperature Hall coefficient which is extrapolated to a sudden jump at $T=0$. 
Since the low-temperature Hall coefficient is related to the FS topology the results 
are interpreted as showing a sudden reordering of the FS at the QCP from a 
{\it large} FS, where the local moment impurity spins are included in the Luttinger 
volume, 
to a {\it small} FS where the impurity spins drop out of the FS 
volume.  This scenario lies at odds with the Hertz-Millis  description of the quantum 
phase transition \cite{Hertz76,Millis93} and has triggered alternative descriptions 
\cite{Si01,Senthil04}.
In this study it is exactly this issue which we investigate using the Kondo lattice as our model 
system:
\begin{equation}
H=  \sum_{ {\mathbf k},\sigma } \epsilon( {\mathbf k}  )  
   c^{\dagger}_{ {\mathbf k}, \sigma} c_{ {\mathbf k}, \sigma } + 
    J \sum_{ \mathbf i }  {\mathbf S}^{c}_{ \mathbf i} \cdot {\mathbf S}^{f}_{ \mathbf i}
\end{equation} 
with $c^{\dagger}_{ {\mathbf k},\sigma}$ creating a conduction electron on an extended orbital with wave vector ${\mathbf k}$ and a z-component of spin $\sigma=\uparrow , \downarrow$. The spin $1/2$ degrees of freedom, coupled via $J$, are represented with the aid of the Pauli spin matrices ${\pmb \sigma}$ by ${\mathbf S}^{c}_{ \mathbf i}=\frac{1}{2} \sum_{s,s'} c^{\dagger}_{ {\mathbf i}, s} {\pmb \sigma}_{s,s'} c_{ {\mathbf i}, s'} $ or the equivalent definition for ${\mathbf S}^{f}_{ \mathbf i}$ using the localized orbital creation operators $f^{\dagger}_{{\mathbf i},\sigma}$.
The KLM forbids charge fluctuations on the $f$-orbitals and as such the constraint of one electron 
per localized orbital must be included. To avoid particle-hole symmetry at half-band filling, 
we  opt for  a dispersion relation:  $ \epsilon(  {\mathbf k} ) =  
-2 t [ \cos(k_x) + \cos(k_y) ] - 2t' [ \cos(k_x + k_y) +  \cos(k_x - k_y) ] $ where $t$ and $t'$ are the first and second nearest neighbor hoping matrix elements respectively. We use $t'/t = -0.3$ throughout 
our calculations. 
\begin{figure}
\includegraphics[width=\columnwidth]{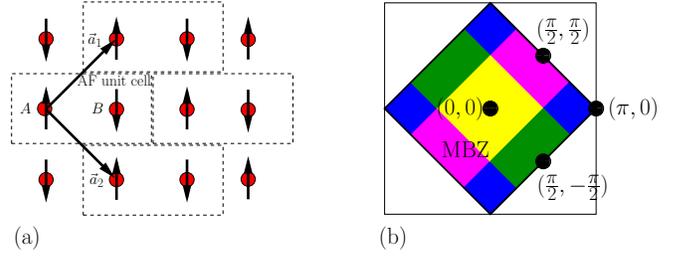}
\caption{(a) Definition of the real space basis vectors, unit cell and orbital indices (b) The reduced Brillouin zone and k-space patching for a cluster with $N_{c}^{AF}=4$}
\label{fig:AFcell} 
\end{figure}

{\it Method -} To solve the above model, we apply the dynamical cluster approximation 
\cite{Hettler00,Maier05} (DCA) with a Hirsch-Fye quantum Monte Carlo (QMC) algorithm 
as cluster solver.
The DCA relies on a $k$-space coarse graining.  The Brillouin zone is divided into $N_{c}$ 
{\it patches} with center $ \pmb K $  and the original $ \pmb k$-vectors are given by: 
$\pmb k = \pmb K + \pmb {\tilde {k}}$.   The DCA boils down to satisfying momentum 
conservation only between patches. As  can be seen within 
a skeleton  expansion,  this approximation yields a  self-energy which is a functional of 
the coarsed-grained  Green function: 
 $\bar{G}(\pmb K, i \omega_{n})=\frac{N_{c}}{N} \sum_{\pmb {\tilde {k}} \in patch} 
 G(\pmb K + \pmb {\tilde {k}} , i \omega_{n})$. With   
$G(\pmb k , i \omega_{n})= (G^{-1}_{0}(\pmb k , i \omega_{n})-\Sigma[\bar{G}
(\pmb K, i \omega_{n})])^{-1}$  we obtain a  self-consistent equation for  $ \bar{G} $. Each 
iteration step   requires  the solution of  $N_c$ impurities   embedded  in a 
fermionic bath  given by: ${\cal G}_{0}^{-1} = \bar{G}^{-1} + \Sigma $.

Our particular implementation of the DCA allows for spontaneous antiferromagnetic 
symmetry breaking which enables us to not only 
localize the magnetic transition but also to carry out simulations within the magnetically 
ordered phase.  The setup of our calculation is shown in Fig.~\ref{fig:AFcell}a.  
Broken lattice symmetries are taken 
into consideration by defining a unit cell of two lattice sites containing, 
in total, two conduction electron orbitals and two localized $f$-electron orbitals.   Within 
this unit cell, the up and down bath Green functions are allowed to take different values  thus 
generating spin symmetry broken solutions. 
This is the smallest cluster, which we denote by $N_{c}^{AF}=1$, with which we can capture 
AF ordering.
Translational symmetry is now only assumed for the new basis vectors 
${\mathbf a}_{1}$ and ${\mathbf a}_{2}$ 
giving the magnetic Brillouin zone (MBZ) in momentum space. 
DCA patching of the MBZ is demonstrated in Fig.~\ref{fig:AFcell}b for a 
cluster of size $N_{c}^{AF}=4$. 
Importantly, the DCA self-consistent equation for the lattice Green function becomes a $4 \times 4$-matrix equation reflecting the four orbitals of the unit cell. 
In this study we use both $N_{c}^{AF}=1$ and $N_{c}^{AF}=4$ clusters. Although, for instance, with $N_{c}^{AF}=1$ the coarse-grained self-energy has no momentum dependence a full $\pmb {k}$-dependency of the lattice Green function enters via the non-interacting single-particle Green function $G_{0}({\pmb k}, i \omega_{n})$, thereby allowing the calculation of a detailed FS regardless of the cluster size.
Our implementation of the  Hirsch-Fye impurity algorithm   
\cite{HirschFye86} to solve the cluster problem follows  precisely the  ideas introduced in 
Ref.~\cite{Capponi00}.  

\begin{figure}
\includegraphics[width=\columnwidth]{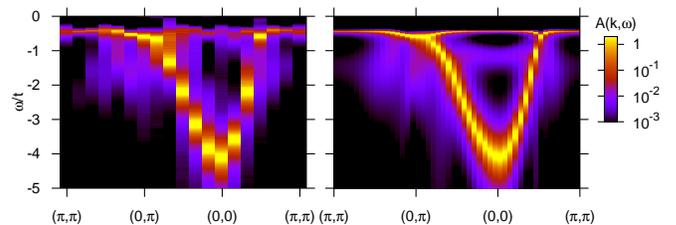}
\caption{(color online) Comparison of the single particle spectral function of the KLM at half-filling
 ($ t'/t = 0$ and $\langle n_c \rangle = 1 $) and with coupling $J/t=1.2$ from 
(left) $12 \times 12$-lattice QMC results as obtained with the 
projective auxiliary field algorithm of Ref. \cite{Capponi00} and 
(right) DCA results with cluster size $N_{c}^{AF} =1$ and $\beta t=40.0$}
\label{half-filling} 
\end{figure}

{\it Results -}  The DCA is justified if the underlying physics is driven by the frequency dependence of 
the self-energy as opposed to its momentum dependence.  To check this  we have carried out 
simulations  at the particle-hole  symmetric point ($ t'/t = 0$ and $\langle n_c \rangle = 1 $) 
and compared the results 
with the  lattice  QMC simulations of Ref.~\cite{Capponi00}.  
Although the DCA at $N_{c}^{AF} =1 $ grossly overestimates the coupling at which 
the  magnetic order-disorder transition occurs,  
it reproduces the single particle spectral 
function both in the paramagnetic (PM) phase and, most importantly, in the AF phase remarkably well. 
This is demonstrated in Fig.~\ref{half-filling} for the AF phase.  Hence, the DCA is able to  
capture the delicate interplay between Kondo screening and  magnetic ordering \cite{Capponi00}.
The sign problem which restricts finite sized lattice QMC studies of the 
KLM to the particle-hole symmetric case is not severe in the DCA approach close to half-filling 
and on cluster sizes up to 16 orbitals. 
\begin{figure}
\begin{center}
\includegraphics[width=\columnwidth]{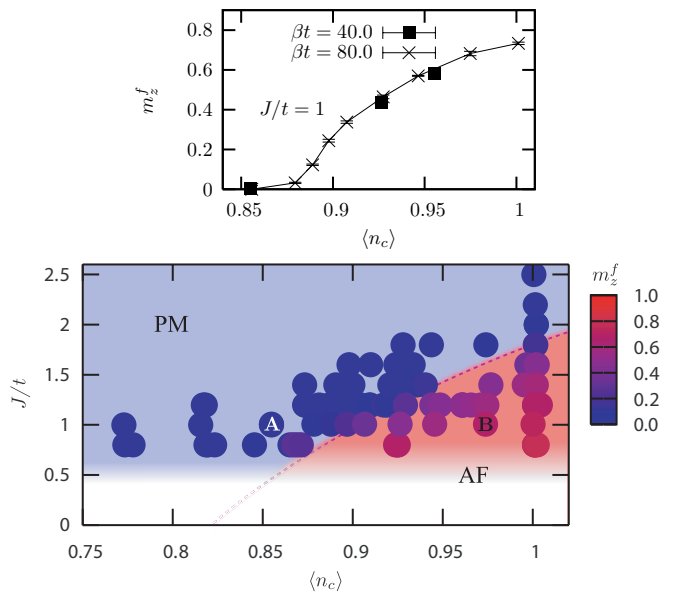}
\end{center}
\caption{(color online) Bottom: Magnetic phase diagram of the hole-doped KLM showing simulation results for the staggered magnetization $m^{f}_{z}$ (color-coded circles) as a function of coupling $J/t$ and conduction electron occupancy $\langle n_{c} \rangle$. Shading of AF and PM regions is a guide to the eye. Here $t'/t = -0.3$ and the calculations are carried out with the $N_{c}^{AF} = 1$ cluster. Top: $m^{f}_{z}$ as a function of $\langle n_{c} \rangle$ at constant coupling $J/t=1$}
\label{Phase} 
\end{figure}
Fig.~\ref{Phase}  maps out the ground state magnetic phase diagram of the KLM as a function of 
coupling $J/t$ and conduction band hole-doping.  Here, we are  interested in 
ground state properties, so the choice for the inverse temperature 
$\beta t$ must be large enough to 
ensure we are below the smallest scale in the problem: the coherence scale and/or the RKKY scale. 
Since the coherence scale decays exponentially with $J/t$ 
we are limited to values of 
$J/t \geq 0.8$.   This restriction arises since the computational time required by the QMC cluster 
solver scales as $(\beta N^{AF}_{c})^3$ \cite{HirschFye86}.  
On general grounds, we expect  the onset of magnetism at small values 
of $J/t$ since in this region the RKKY scale set by  $J^{2} \chi( {\mathbf q}, \omega = 0)$ 
dominates over the  
Kondo scale, $T_K \sim e^{-t/J} $.  Here, $\chi( {\mathbf q}, \omega = 0)$  corresponds to 
the spin susceptibility of the conduction 
electrons. It is interesting to note that within the DCA approximation on cluster sizes up to 
$N_c^{AF} = 4 $ the  value of $J/t$ at which antiferromagnetism sets in at half-band filling, 
$n_c = 1$,  is  only 
slightly affected  by the choice of $t'/t$.  On the other hand, the charge gap, which is known to 
scale with $J/t$  in the particle-hole 
symmetric case ($t'/t = 0$) \cite{Capponi00}, is very much suppressed  away from the particle-hole symmetric point. In 
particular, at $t'/t = -0.3 $ our results are consistent with an exponential scaling of the charge gap
\cite{Martin07}. 
Upon doping we observe a magnetic metallic state which as a function of decreasing coupling $J/t$ progressively dominates 
the phase diagram. Finally, within our numerical accuracy, the staggered magnetization
vanishes smoothly at the magnetic order-disorder transition thus lending support to a continuous 
transition (see Fig.~\ref{Phase}, top panel).
\begin{figure}
\begin{center}
\includegraphics[width=\columnwidth]{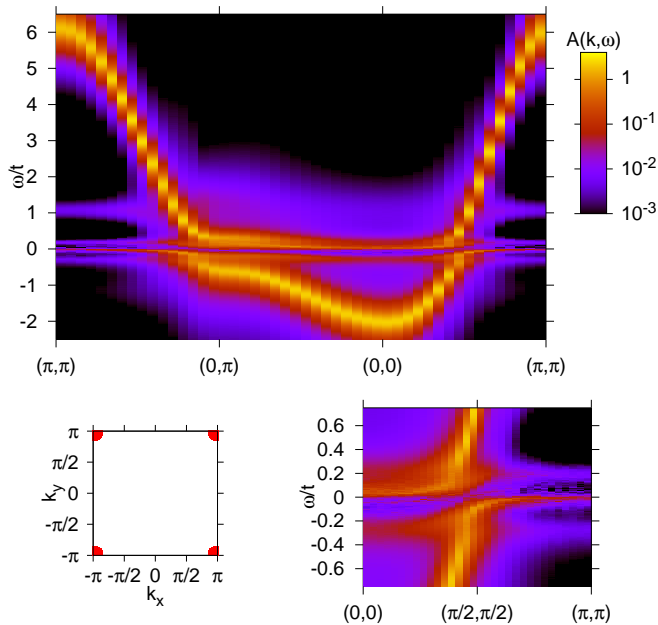}
\end{center}
\caption{(color online) DCA single particle spectrum for simulation A (see Fig.~\ref{Phase}) in the PM region ($J/t=1.0$, $\langle n_{c} \rangle = 0.855$ and $\beta t=40.0$). The bottom right plot is a close-up of the main plot for energies around the Fermi-energy along the path $(0,0)$ to $(\pi,\pi)$. The bottom left plot shows the resultant Fermi-surface, plotted here via a large-$N$ mean-field modeling of the DCA data (see text)}
\label{DCA_PM} 
\end{figure}
\begin{figure}
\begin{center}
\includegraphics[width=\columnwidth]{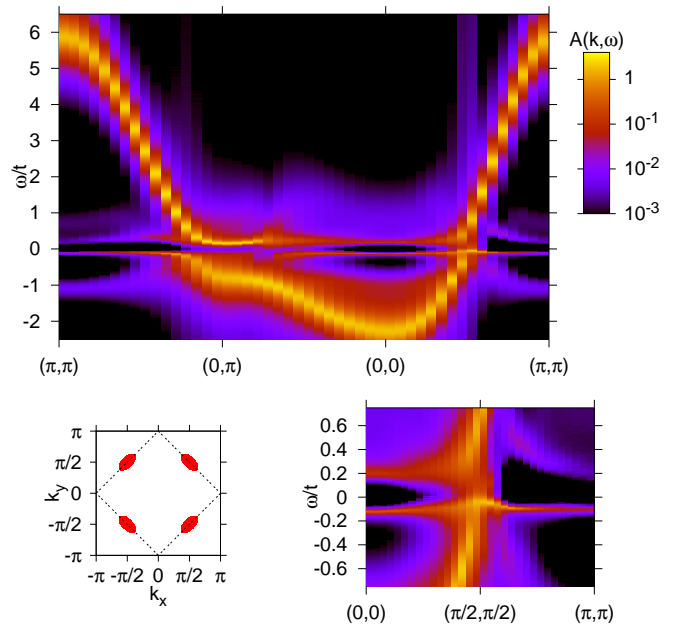}
\end{center}
\caption{(color online) DCA result for the single particle spectrum with $N_{c}^{AF}=4$, $J/t=1.0$, $\langle n_{c} \rangle = 0.977$ and $\beta t=40.0$ in the AF region (point B in Fig.~\ref{Phase}). The bottom right plot is a close-up of the main plot for energies around the Fermi-energy along the path $(0,0)$ to $(\pi,\pi)$. The bottom left plot shows the resultant Fermi-surface plotted here by fitting to the DCA spectrum using the mean-field form of Eq.~\ref{eqn:KLMmeanfield}. }
\label{DCA_AF} 
\end{figure}

Having established  the magnetic phase diagram, we turn to our primary concern, namely 
the evolution of the FS across the magnetic order-disorder transition. Figs.~\ref{DCA_PM} and 
\ref{DCA_AF} plot the single particle spectral function, 
$A(\pmb k,\omega)  = - {\rm Im} G_{cc}^{ret}(\pmb k, \omega) $ in the PM and AF metallic 
phases corresponding to the points A and B respectively in the phase diagram (Fig.~\ref{Phase}). 
Analytic continuation to real frequencies was achieved with a  
stochastic analytic continuation method \cite{Sandvik98,Beach04a}.

The DCA spectral function in the PM phase (Fig.~\ref{DCA_PM}) is characterized by an extremely flat heavy fermion band crossing the FS in the vicinity of $ {\mathbf k} = (\pi,\pi) $ and equivalent points. In the AF phase (Fig.~\ref{DCA_AF}), again at $J/t = 1.0 $ but with $N_c^{AF} = 4$ and in close proximity to half-band filling the heavy fermion band centered around $(\pi,\pi)$ is still present but has dropped 
below the Fermi energy and hole pockets centered around the wave vector ${\mathbf k} = (\pi/2,\pi/2) $ arise. 
This result is further supported  by calculations at  $J/t = 0.8 $, $\beta t = 80$, closer to the 
magnetic transition ($n_c=0.92$) but on smaller cluster sizes ($N_c^{AF} = 1$). 
The continued presence of heavy bands around $(\pi,\pi)$, although shifted to lower energies, 
indicates that partial Kondo screening still occurs in the magnetic phase. 
The main result of our spectral analysis however is the topological change of the FS when 
comparing the PM and AF phases.  

{\it Interpretation -} As an aid to interpreting our DCA results we draw on the  mean-field approximation of
Zhang and Yu \cite{Zhang00b} which introduces the order parameters 
$\langle f_{\mathbf i ,\uparrow}^{\dagger} f_{\mathbf i ,\uparrow}  -  f_{\mathbf i ,\downarrow}^{\dagger} f_{\mathbf i ,\downarrow} \rangle=m_{f} 
e^{i \mathbf Q \mathbf i}$, $\langle c_{\mathbf i ,\uparrow}^{\dagger} c_{\mathbf i ,\uparrow}  -  c_{\mathbf i ,\downarrow}^{\dagger} 
c_{\mathbf i ,\downarrow} \rangle = -m_{c} e^{i \mathbf Q \mathbf i}$, with $Q=(\pi, \pi )$, and 
$\langle f_{\mathbf i ,\uparrow}^{\dagger} c_{\mathbf i ,\uparrow}  +  c_{\mathbf i ,\downarrow}^{\dagger} f_{\mathbf i ,\downarrow} \rangle = 
\langle f_{\mathbf i ,\downarrow}^{\dagger} c_{\mathbf i ,\downarrow}  +  c_{\mathbf i ,\uparrow}^{\dagger} f_{\mathbf i ,\uparrow} \rangle = -V$.
$m_{f}$ and $m_{c}$ are the staggered magnetizations of the impurity spins and conduction electrons respectively and $V$ is a hybridization order parameter which mimics the screening of the impurity spins.  
In this mean-field approach, which we stress is used here purely as a tool to aid interpretation of our DCA results, the constraint of unit occupation on the localized $f$-electron orbital is taken into account only on average by means of a Lagrange multiplier $\lambda$, as opposed to the exact single occupancy enforced within the DCA-QMC cluster solver.

The mean-field Hamiltonian, given by  
\begin{eqnarray}
& &\tilde{H}=\sum_{\mathbf k \in MBZ \sigma} 
\left(
\begin{array}{cccc}
c_{\mathbf k \sigma}^{\dagger} &
c_{\mathbf k + \mathbf Q \sigma}^{\dagger} &
f_{\mathbf k \sigma}^{\dagger} &
f_{\mathbf k + \mathbf Q \sigma}^{\dagger}
\end{array}
\right)\times \label{eqn:KLMmeanfield} \\
& & \left(
\begin{array}{cccc}
\epsilon_{\mathbf k} -\mu     & \frac{J m_{f} \sigma}{4}        & \frac{J V}{2}               & 0                           \\
\frac{J m_{f} \sigma}{4}   & \epsilon_{\mathbf k + \mathbf Q} -\mu &       0                     & \frac{J V}{2}               \\
\frac{J V}{2}              &          0                      &   \lambda                   & -\frac{J m_{c} \sigma}{4}   \\
         0                 &       \frac{J V}{2}             &  -\frac{J m_{c} \sigma}{4}  &   \lambda
\end{array}
\right)
\left(
\begin{array}{c}
c_{\mathbf k \sigma} \\
c_{\mathbf k + \mathbf Q \sigma} \\
f_{\mathbf k \sigma}\\
f_{\mathbf k + \mathbf Q \sigma}
\end{array}
\right) \nonumber
\end{eqnarray}
results in a four-band energy dispersion relation, $E_{n}(\mathbf k)$, in the MBZ.

The low temperature  features of the PM phase are well understood in this mean-field framework 
by setting $ m_f = m_c  = 0$ in Eq.~(\ref{eqn:KLMmeanfield}) which recovers the generic 
hybridized  band structure of the large-$N$ approximation,
$E_{\pm}(\mathbf k) = \frac{\epsilon_{\mathbf k}-\mu+\lambda}{2} \pm \frac{1}{2} \left( \left( \epsilon_{\mathbf k} -\mu-\lambda \right)^{2} + (J V)^{2} \right)^{\frac{1}{2}}$ with heavy bands crossing the Fermi energy in the vicinity of ${\mathbf k} =(\pi,\pi)$ and equivalent points in excellent agreement with the DCA result.
In this mean-field modelling of the PM phase the Luttinger volume is given by 
$V^{PM}_L/V_{BZ}  = (n_f + n_c)/2$, counting both conduction band electrons and 
impurity spins. 
Since the mean-field model accounts very well for the DCA spectral function in the PM phase, with the exception of the spectral weight around ${\mathbf k} = (0,0) $ and $ \omega/t \simeq -1/4 $ 
\footnote{We note that this feature is present both in the DCA on clusters up to 
16 orbitals as well as in the DMFT approach. Hence, it cannot originate from magnetic fluctuations. A detailed 
understanding of this feature is presently under investigation }, we attribute the DCA result with this same {\it large} Luttinger volume.

In the AF phase the DCA spectrum can again be well accounted for with the mean-field Hamiltonian of 
Eq.~(\ref{eqn:KLMmeanfield}) but now setting non-vanishing staggered magnetizations, 
equal to the QMC measured observables, and using a non-zero hybridization value of $V=0.3$ as 
fit parameter. 
In this fit, one band drops completely below the Fermi energy which itself is crossed only by the second band $E_{2}(\mathbf k)$ in the vicinity of ${\mathbf k} = (\pi/2,\pi/2)$.
In this case the Luttinger volume is given by $V^{AF}_L/V_{MBZ}  = n_c$ and the topology of the FS is that of a spin-density wave approximation  ($V = 0$, $m_f =1$,$m_c \neq 0$ in Eq.~(\ref{eqn:KLMmeanfield})) where the 
$f$-electrons are frozen and do not participate in the Luttinger sum rule. Hence,  we coin the FS, obtained from the DCA spectrum, as {\it small}.

We note that similar results have recently been observed in  
a  variational Monte-Carlo approach to the KLM \cite{Watanabe07}. 
A major difference however is that the magnetic phase transition separating topologically 
different Fermi surfaces is  of first order in the variational approach \cite{Watanabe07}.
In contrast, the DCA calculation presented here 
supports  a continuous transition in accordance with experimental findings.

{\it Conclusions -} We have presented large scale DCA calculations of the Kondo lattice model, and 
mapped  out the magnetic phase diagram in the $J/t$ versus doping plane.  We have tested the
approximation and seen remarkable agreement with lattice QMC methods  with respect to the single 
particle spectral function.  In particular, the approximation captures the 
delicate interplay between Kondo screening and  magnetic ordering. 
Across the magnetic phase transition, the data shows a change in the topology of the 
FS from a  large FS in the  PM state to a small FS with hole-pockets centered around  
${\mathbf k} = (\pi/2,\pi/2)$ 
in the AF phase.  This change in topology of the FS is not linked to the breakdown of Kondo screening, 
and is consistent with a jump in the Hall coefficient as observed experimentally \cite{Paschen04}. 
This jump in the Hall coefficient is not linked to a first order transition since within  our
numerical accuracy the  staggered magnetization  is a continuous function of the control parameter 
driving the quantum phase transition.

We would like to thank  the  Forschungszentrum  J\"ulich  for generous allocation of CPU time on 
the IBM Blue Gene/L and the DFG for financial support. We thank K. Beach, S. Capponi, S. Hochkeppel, 
T. C. Lang, T. Pruschke and M. Vojta for conversations.

\end{document}